\documentclass[]{mn2e}

\usepackage{graphics}
\usepackage{graphicx}
\usepackage{amsxtra}
\usepackage{array,longtable}

\usepackage{latexsym}
\usepackage{amssymb}

\usepackage{ulem}

\usepackage{color}
\title[Malin\,1: interacting galaxy pair?]{Malin\,1: interacting galaxy
pair?\thanks{Partly based on observations collected with the 6m
telescope at the Special Astrophysical Observatory (SAO) of the
Russian Academy of Sciences (RAS).}}
\author[V.P. Reshetnikov, A.V. Moiseev and N. Ya. Sotnikova]{V. P.
Reshetnikov$^1$\thanks{E-mail: resh@astro.spbu.ru}, A. V. Moiseev$^2$ and
N. Ya. Sotnikova$^1$
\\
$^1$St.Petersburg State University, Universitetskij pr. 28, 198504
St.Petersburg, Stary Peterhof, Russia \\
and Isaac Newton Institute of Chile, St.Petersburg Branch\\
$^2$Special Astrophysical Observatory, N.Arkhyz, Karachaevo-Cherkesia,
369167, Russia }

\begin{document}

\date{Accepted 2010 May 27. Received 2010 May 27; in original form 2010 February 11}


\maketitle


\begin{abstract}

Malin\,1 is a unique, extraordinarily large low surface brightness
galaxy. The structure and the origins of the galaxy are poorly
understood. The reason for such a situation is an absence of detailed
observational data, especially, of high-resolution kinematics.
In this Letter we study the stellar kinematics of the inner part
($r \leq 15$\,kpc) of Malin\,1.

We present spectroscopic arguments in favour of a small galaxy --
Malin\,1B -- being a companion probably interacting with the main galaxy --
Malin\,1. This object is clearly seen in many published images
of Malin\,1 but is not mentioned in any astronomical databases.
Malin\,1B is located at the projected distance of
14 kpc from the Malin\,1's nucleus and has small -- 65$\pm$16 km/s --
relative velocity, which we determined for the first time.
We suggest that ongoing interaction with Malin\,1B can explain main
morphological features of the Malin\,1's central region -- two-armed
spiral structure, a bar, and an external one-armed spiral pattern.

We also investigated the large scale environment of Malin\,1
and postulate that the galaxy SDSS\,J123708.91+142253.2 might be
responsible for the formation of extended low-surface brightness 
envelope by means
of head-on collision with Malin\,1 (in the framework
of collision scenario proposed by Mapelli et al. 2008).
To test the collisional origins of Malin\,1 global structure, more
observational data and new numerical models are needed.

\end{abstract}


\section{Introduction}

Malin\,1 is one of the most unusual galaxies known to date.
The galaxy was accidentally discovered in the course of a systematic
survey of the Virgo cluster region designed to detect extremely low-surface
brightness (LSB) galaxies (Bothun et al. 1987).
According to observational characteristics, Malin\,1 remains unique
among other known galaxies.

Radial extent of its stellar disc in the $V$ passband reaches out
$\approx$60$\arcsec$ or $\approx$90 kpc
(adopting distance $D_L$=366 Mpc and scale 1.51 kpc/1$\arcsec$ from the
NASA/IPAC Extragalactic Database (NED))
and the disc scale length is about 45$\arcsec$ or 68 kpc (Bothun et al. 1987).
Deep $R$-band data show even larger radial size of the disc --
about $80\arcsec$ or 120 kpc -- and somewhat smaller scale length
value --  33$\arcsec$ or 50 kpc (Moore \& Parker 2006).
Therefore, Malin\,1 possesses the largest stellar disc of any known spiral
galaxy. The disc is of extremly low surface brightness --
extrapolated central surface brightness
is $\mu_0(V) \approx 25\fm5$/$\square\arcsec$ (Bothun et al. 1987),
$\mu_0(R) \approx 24\fm7$/$\square\arcsec$ (Moore \& Parker 2006).
Malin\,1 has a very low brightness, but due to its enormous size
the galaxy's total optical luminosity is high: $M_V \approx -22\fm9$ (Pickering et al. 1997).
Based on these characteristics, Malin\,1 is often considered as
a prototypical giant low surface brightness (LSB) galaxy.

Malin\,1 is among the most gas-rich galaxies known to date:
M(HI)$\approx 7 \times 10^{10}$\,M$_{\odot}$ (Pickering et al. 1997)
or $\approx 5 \times 10^{10}$\,M$_{\odot}$ (Matthews, van Driel \&
Monnier-Ragaigne 2001).
The HI disc of the galaxy demonstrates strong non-circular motions and
is strongly warped (Pickering et al. 1997; Lelli, Fraternali \& Sancisi 2010).

Recent analysis of a Hubble Space Telescope $I$-band image
suggests that Malin\,1 has a normal barred inner spiral disc embedded
in a huge diffuse LSB envelope (Barth 2007).
Based on previously published photometric characteristics and HI
kinematics, Seigar (2008)
derived a possible mass profile for Malin\,1.
He concluded that the galaxy is baryon dominated in the centre
(out to $\sim$10 kpc) and, probably, has parameters typical of
normal galaxies (of course, excluding giant low-surface brightness
envelope). Recently Lelli et al. (2010) (see also Sancisi \& Fraternali 2007)
presented the similar results
using their re-analysis existing HI data. The models by Seigar (2008)
and Lelli et al. (2010) are based on published HI observation by Pickering
et al. (1997). But these HI data are strongly affected by beam
smearing and, as a result, no reliable kinematics inside central 15
kpc are currently available.

In order to investigate stellar kinematics of Malin\,1, we have performed
spectroscopic study of the galaxy with the Russian 6-m telescope.
Detailed discussion and results of modeling will be published
later. In this Letter, we present the most surprising result
of our observations -- the discovery of the galaxy (Malin\,1B) that
probably interacts with Malin\,1.

\section[]{Spectral observations}

The spectroscopic data were obtained with the 6-m telescope  of the
Special Astrophysical Observatory of Russian Academy of Sciences (SAO RAS)
on March 30/31 2009. We observed Malin\,1 with the multi-mode
focal reducer  SCORPIO (Afanasiev \& Moiseev 2005)
in the long-slit mode with the spectral resolution about $2.5$\AA\,
in the wavelength range 5500-6550\AA\, that corresponds to 5080-6050\AA\,
in the galaxy rest frame. The studied spectral range contains
numerous  absorption features (MgI, Fe, etc.) produced by old stellar
population. The possible contribution of   weak  emission lines was insignificant.
The slit, which length is about 6 arcmin, has
been placed on the galaxy nucleus at P.A.=55$^\circ$ (see Fig.~1).
The slit width was $1\arcsec$ with the spatial sampling
$0.35\arcsec/px$. We took in total 7200 sec exposure under atmospheric
seeing $2\arcsec$.

For data reduction and analysis we used programmes and algorithms
briefly described in Zasov et al. (2008). The line-of-sight velocity and
stellar velocity dispersion profiles for the stellar components have been
estimated by cross-correlating galactic spectra binned along the slit
with a template star HD~10380 (K3~III) spectra from the library MILES
(Sanchez-Blazquez et al. 2006). We applied 2-3 pixels  binning along the slit to
provide a sufficient signal-to-noise ratio, the resulting sampling was
0\farcs70 for the inner part of the galaxy and 1\farcs05 for several
external points.

The results are presented in Fig.~2. At $r\approx9\arcsec$ the slit
crosses the nucleus of small companion galaxy (Malin\,1B). Fig.~2 shows
that this
region is kinematically decoupled from the Malin\,1's rotation curve: it
has an invert radial velocity gradient and a distinct peak at the
velocity dispersion profile.

\begin{figure}
\centering
\includegraphics[width=8.8cm, clip=]{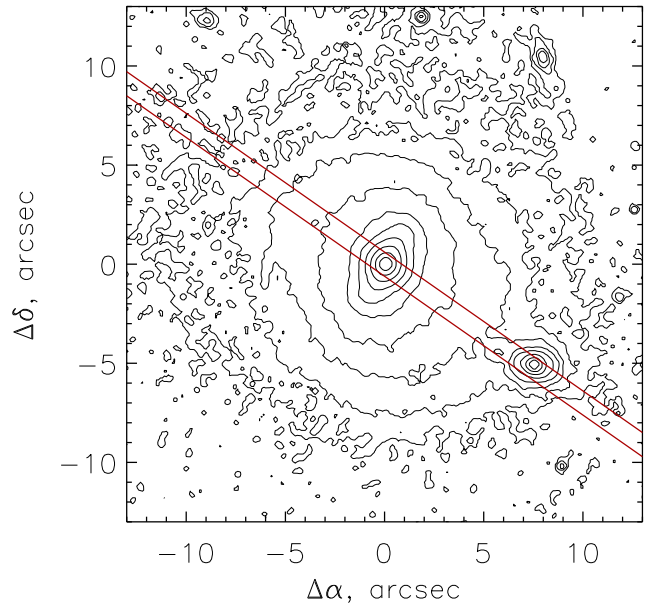}
\caption{$I$-band contour map of the central part of Malin~1.
The faintest contour is about 23\fm7/$\square\arcsec$, the isophotes step --
0\fm75/$\square\arcsec$. The red lines mark position of the  long-slit
(1 arcsec in width). The F814W WFPC2 image was extracted from the Hubble
Legacy Archive. }
\end{figure}

\begin{figure}
\centering
\includegraphics[width=8.8cm, clip=]{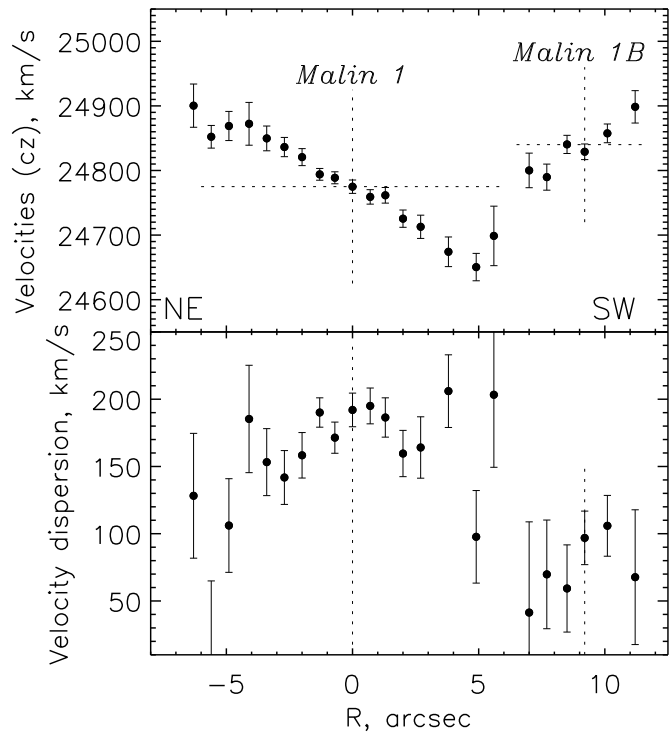}
\caption{The line-of-sight velocity (top) and velocity dispersion
(bottom) profiles of stars at P.A.=55$^\circ$. Dotted lines mark
photometric positions of Malin\,1 (main galaxy) and Malin\,1B nuclei
and their systemic velocities.}
\end{figure}

\section[]{Results and discussion}

\subsection[]{Characteristics of Malin\,1}

From our spectrum we determined the following parameters of
the main galaxy (Malin\,1): the heliocentric systemic velocity
V$_\mathrm{sys}$=24775$\pm$10 km/s
and the central velocity dispersion
$\sigma_0$=192$\pm$13 km/s. Both values are in good agreement with
previously published results (Table~1).

The companion galaxy -- Malin\,1B -- has coordinates
$\alpha$(2000)=12$^h$36$^m$58.$^s$89 and
$\delta$(2000)=+14$^\circ$19\arcmin43\farcs9 and is located about
9$\arcsec$ SW from the main galaxy nucleus (Fig.~1).
The systemic radial velocity of the companion is V$_{sys}$=24840$\pm$12 km/s,
the velocity difference with Malin\,1 is 65$\pm$16 km/s,
the projected distance is 14 kpc.

Outer elliptical isophotes of Malin\,1B are disturbed and elongated
to the North-North-East (Fig.~3). The mean axial ratio the isophotes is
$b/a = 0.76 \pm 0.07$, the position angle
of the major axis is P.A.=79$^\circ$ $\pm$ 3$^\circ$, and the major
axis is about 3$\arcsec$ or 4.5 kpc.

\begin{figure}
\centering
\includegraphics[width=10.5cm, angle=-90, clip=]{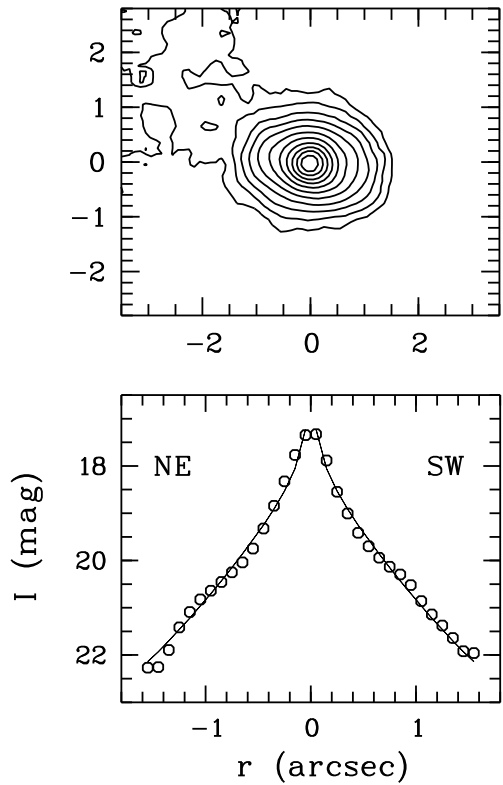}
\caption{Top: contour map of Malin~1B from the HST $I$-band image.
Orientation of the figure is the same as in Fig.~1. Isophotes are separated by
factor 1.5, faintest contour is 22\fm3/$\square\arcsec$. Axes are in arcsec.
Bottom: photometric profile for Malin\,1B along the apparent major
axis (dots). The solid line represents approximation of the profile
by standard two-component model (de Vaucouleurs' bulge and exponential disk).}
\end{figure}

The surface brightness distribution of Malin\,1B looks usual for
early-type spiral galaxies (Fig.~3). In the framework of two-component model
the bulge-to-disk ratio is $\approx1$. This value is
typical for the S0-Sa galaxies (e.g. Simien \& de Vaucouleurs 1986).

The observed luminosity of Malin\,1B measured within
$\mu(I)=23^m/\Box\arcsec$ is about 1/10 of the Malin\,1 luminosity
within the same isophote. 
The total luminosity of the companion
$M_I \approx -20$ or $M_V \approx -18.7$
(assuming $V-I=1.3$ for an S0 galaxy). 

The observed central velocity dispersion of Malin\,1B is
$\sigma_0$=97$\pm$20 km/s and the galaxy parameters satisfy
the Faber--Jackson relation.

The ratio $V_{max}/\sigma_0 > 0.7$ is typical for the
rotationally-supported
disc galaxies, as it follows, for instance, from Binney's diagnostic
$(v/\sigma-\epsilon)$ diagram (e.g. Kormendy 1993).

\begin{table*}
\caption{General characteristics of the galaxies}
\begin{tabular}{lllll}
\hline
\hline
Parameter &  Malin\,1 & Malin\,1B  & SDSS\,J123708.91+142253.2 & Ref. \\
\hline
$\alpha(2000)$     & 12$^h$36$^m$59.$^s$36 & 12$^h$36$^m$58.$^s$89 &12$^h$37$^m$08.$^s$92
\\
$\delta(2000)$     & 14$^\circ$19\arcmin49\farcs4 & 14$^\circ$19\arcmin43\farcs9 & 14$^\circ$22\arcmin53\farcs3
               \\
                   &                  &    &  & \\
Projected separation from Malin\,1    & 0   & 9$''$ (14 kpc) & 3\farcm855 (350 kpc) \\
                   &                  &    &  & \\
Apparent magnitude & 17.6 (g) &       & 18.6 (g) & SDSS \\
                   &                  &    &  & \\
Heliocentric systemic velocity$^*$ (km~s$^{-1}$)& 24775$\pm$10 & 24840$\pm$12 & & present work \\
                                                & 24800$\pm$17 &      & 24907$\pm$27  & SDSS \\
                                                & 24819$\pm$65 &      &               & SDSS \\
                                                & 24767$\pm$4  &  &  & Lelli et al. (2010) \\
                                                & 24784$\pm$15 &  &  & Matthews et al. (2001) \\
                                            & 24755$\pm$10 &  &  & Pickering et al. (1997) \\
                                & 24705, 24745 &  & & Impey \& Bothun (1989) \\
                                                & 24750$\pm$10 &  & & Bothun et al. (1987) \\
                   &                  &    &  & \\
Central velocity dispersion (km~s$^{-1}$) & 192$\pm$13 & 65$\pm$16 & & present work\\
                                          & 196$\pm$15 &   &  & Barth (2007) \\
\hline \\
\end{tabular}

$^*$ -- conventional radial velocity obtained as $cz$
\end{table*}

\subsection{Malin\,1 as an interacting galaxy}

Fig.~4 shows the large-scale spatial environment of Malin\,1.
As one can see, the galaxy is located in a relatively low-density region,
near the end of an elongated structure, that is probably a filament of the
large scale structure (LSS). Therefore, Malin\,1 is within the environment
typical for LSB galaxies.
For example, Rosenbaum \& Bomans (2004) and Rosenbaum et al. (2009)
concluded that the LSB galaxies appear to favour the
edges of filaments.
This conclusion supports the idea that LSB galaxies were formed in the voids
of the LSS without many galaxy interactions. Only in this case
the low-density primordial fluctuations could survive. Then the galaxy
would have migrated to the edges of the filaments due to gravitational infall.

\begin{figure}
\centering
\includegraphics[width=8.cm, angle=-90, clip=]{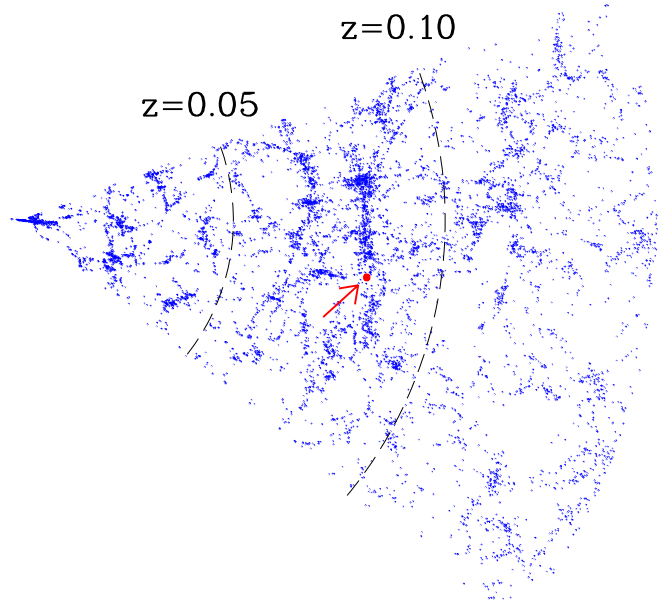}
\caption{Distribution of galaxies within a right ascension range of
$160^{\circ} \leq \alpha(2000) \leq 220^{\circ}$ (from top to bottom)
and a redshift of $z\leq0.15$ in a polar plot according to the NED database.
Malin\,1 position is indicated by red arrow. The declination range of
$13.1^{\circ} \leq \delta(2000) \leq 15.6^{\circ}$ is projected
onto the plane.}
\end{figure}

There is another point of view about the origins of LSB galaxies.
Some authors suggested a possible link between giant low-surface brightness
galaxies
and ring galaxies formed as a result of a head-on collision with a massive
flyby intruder (e.g., Mapelli et al. 2008). Smoothed shell-like structures at
large radii can be also produced by mergers between a disc-like
galaxy and elliptical one (Pierani et al. 2010).

The collisional scenario is very attractive because Malin\,1 has
a normal inner stellar disc and a normal bulge (Barth 2007).
Moreover, its extended stellar and gaseous discs can be considered as
collisional rings at the late stage of their evolution. Mapelli et al. (2008)
presented a model that matches in general the global photometry of Malin\,1.
Unfortunately, such a model needs the intruder with mass that is comparable
with the total mass of the main galaxy or only a few times less.
Malin\,1B looks too small and unsufficiently massive to produce an extended
ring-like structure.
It could have been more massive originally but has been tidally stripped during the
interaction with Malin\,1. But to our mind,
its position and relative velocity
favour a close orbit rather than a flyby at escape speed that is commonly
preassumed when one considers a collisional origin of ring-like galaxies
(Lynds \& Toomre 1976; Hernquist \& Weil 1993; Horellou \& Combes 2001).

Nevertheless Malin\,1B could cause the appearance of a single dusty spiral
arm in Malin\,1 that has been recently revealed (Moore \& Parker 2006).
This type of structure is often considered as the result of an interaction
with a satellite on a retrogade orbit (Athanassoula 1978).

Malin\,1B could also produce two-armed structure and trigger
a bar instability in the inner disc of Malin\,1.
The inner spiral structure is definitely seen in the Hubble Space
Telescope $I$-band image (Figs.~1,2 of Barth 2007) and the bar can
be clearly recognized by analysing the surface brightness distribution
of Malin\,1 (Fig.~3 of Barth 2007). But we definitely need a more
massive companion to support the scenario by Mapelli et al. (2008)

The nearest bright galaxy with known redshift is SDSS\,J123708.91+142253.2
with $V_\mathrm{hel} = 24907 \pm 27$ (Table~1). The galaxy is located at
the projected distance of 3.9 arcmin or 350 kpc far from Malin\,1 and has an
apparent magnitude $g(\mathrm{SDSS})=18.6$.
The velocity difference between this
galaxy and Malin\,1 is rather small --- 132$\pm$29 km/s.

We can use parameters of
the collisional model presented by Mapelli et al. (2008) to check the
galaxy SDSS\,J123708.91+142253.2 as a candidate for being a possible intruder.
The ratio of the Malin\,1 and  ``intruder'' optical luminosities
(that is about $1/2.5$ in the $g$-band) is in agreement with the
value of the mass ratio of target and intruder galaxies
in the model under discussion.

The collisional model (Mapelli et al. 2008) suggests an impact
at an inclination angle of $\sim 14^\circ$ with respect to the angular
momentum axis of the target (Malin\,1 in our case).
According to Moore \& Parker (2006) the inclination angle of
a stellar disc of Malin\,1 is $\sim 45^\circ$. Then the direction of
intruder motion would have an inclination angle to the line of sight between
$\theta \sim 30^\circ$ and $\theta \sim 60^\circ$. Let us choose the
largest value of the angle for further estimates.
The initial position and the velocity (with respect to the centre-of-mass
of the progenitor of the ring galaxy) for the intruder adopted
by Mapelli et al. (2008)
are $r_\mathrm{} \approx 33$ kpc, $v_\mathrm{ini} \approx 900$ km/s.
It implies that the intruder radial velocity would be of
$v = v_\mathrm{ini} \cos{\theta} \sqrt{r_\mathrm{ini} / r} \sim 130$~km/s
at the projected distance $r_\mathrm{proj} = r \sin{\theta} = 350$ kpc.
Both estimates mean that it
takes the intruder about 1~Gyr ($t \approx \frac{2}{3} r / v$) to get
there, providing a very small impact parameter and a flyby
as suggested by Mapelli et al. (2008).

Our simple velocity estimate for the intruder 
is in agreement with the data presented
here. Moreover, current location of the galaxy SDSS\,J123708.91+142253.2
maintains the idea for the closest passage to happen 1~Gyr ago
that is also in accordance with the model by Mapelli et al. (2008).
But this otherwise attractive scenario has one serious drawback.

The model by Mapelli et al. (2008) predicts the existence
of strong non-circular motions that could be associated with the expansion
velocity of the old ring and/or with the fallback of a part of the
ejected matter towards the centre. The strength of non-circular motions
can be quantified by the ratio of radial and tangential velocities
of gas. Simulations show that there is a region in the outer part of the
galaxy ($\sim$70--80 kpc far from the center)
where radial velocities are
comparable with tangential ones (Fig.~12 of Mapelli et al. 2008).
However, the analysis of HI data
(to be discussed in a forthcoming paper) shows that possible radial
motions are smaller (see also Lelli et al. 2010). Namely, the radial component of
non-circular gas motions in the galaxy outskirts have a maximal values
about $0.3-0.4$ of rotation velocities. Moreover, the
azimuthal distribution of non-circular motions is more complex than
simple inflow/expansion pattern predicted by numerical simulations.

In any case, the structure of Malin\,1 seems  more complex than predicted
from Mapelli et al. (2008) model because the real situation should include
two events: a possible strong head-on collision with a massive companion
(SDSS\,J123708.91+142253.2) and a current interaction with
smaller Malin\,1B galaxy.

Thus, we do not reject the collisional model of Malin\,1 but it
definitely needs to be refined, as far as, new observational data are
urgently needed.

\section{Conclusions}

In this letter we present the results of spectroscopic observations of
the galaxy Malin\,1 and its newly discovered companion Malin\,1B.
Our main conclusions are

\begin{enumerate}

\item Malin\,1 is currently undergoing a minor merger with a relatively 
small companion
galaxy. This galaxy is at the projected distance of 14 kpc and shows
a small velocity difference with Malin\,1 systemic velocity
($\Delta$V=65 km/s).

\item Accreting companion might be responsible for the main morphological
features in the central part of Malin\,1. It could produce a two-armed
inner structure and trigger the bar instability (Fig.~1). Also,
one can relate this companion with one-armed spiral pattern in
Malin\,1 disc (see Fig.~4 in Moore \& Parker 2006).

\item Malin\,1 is in a relatively 
low-density large scale spatial environment that is
typical for LSB galaxies.

\item We discuss the possible origins of Malin\,1 global structure due
to a bygone head-on collision with a massive intruder (mechanism proposed
by Mapelli et al. 2008). Available data do not contradict this scenario
(for instance, we even identified possible intruder galaxy
SDSS\,J123708.91+142253.2) but more
detailed simulations and new observational data are needed
for definite conclusions.

\end{enumerate}

\section*{Acknowledgments}

The research is partly based on observations made with the NASA/ESA Hubble Space
Telescope, and obtained from the Hubble Legacy Archive, which is a
collaboration between the Space Telescope Science Institute (STScI/NASA), the Space  Telescope European Coordinating Facility (ST-ECF/ESA) and the  Canadian Astronomy Data Centre (CADC/NRC/CSA).
Funding for the Sloan Digital Sky Survey (SDSS) and SDSS-II has been
provided by the Alfred P. Sloan Foundation, the Participating Institutions,
the National Science Foundation, the U.S. Department of Energy, the National
Aeronautics and Space Administration, the Japanese Monbukagakusho,
and the Max Planck Society, and the Higher Education Funding Council
for England. The SDSS Web site is $http://www.sdss.org/$.
A.V.M. thanks the grant of the Russian Foundation for Basic
Researches number 09-02-00870 and also  the Dynasty Fund. 

We thank an anonymous referee for helpful comments on the paper.


\end{document}